\documentclass[prb,twocolumn,superscriptaddress,altaffillsymbol,amsmath,nofootinbib,longbibliography]{revtex4-1}

\usepackage{ifthen}
\usepackage{epsfig}
\usepackage{booktabs} 
\usepackage{natbib}
\usepackage{wasysym}
\usepackage[many]{tcolorbox}   

\usepackage{slashed} 
\usepackage{bbm}
\usepackage{mathrsfs}
\usepackage{amssymb}
\usepackage{dsfont}
\usepackage[export]{adjustbox}

\usepackage{color}

\allowdisplaybreaks

\newcommand{\beqra}{\begin{eqnarray}}
\newcommand{\eeqra}{\end{eqnarray}}
\newcommand{\beq}{\begin{equation}}
\newcommand{\eeq}{\end{equation}}

\renewcommand{\epsilon}{\varepsilon}

\renewcommand{\vec}[1]{\mathbf{#1}}
\renewcommand{\bar}{\overline}

\tcbset{
    sharp corners,
    colback = white,
    before skip = 0.2cm,    
    after skip = 0.5cm      
}                           

\newtcolorbox{boxA}{
   boxrule = 1.5pt,    
   colframe = black  }

\begin{document}

\title{\boldmath Why are there so few non-altermagnetic antiferromagnets?}

\author{Nicola A. Spaldin}
\email{nicola.spaldin@mat.ethz.ch}
\affiliation{Department of Materials, ETH Zurich, CH-8093 Z\"urich, Switzerland}
\email{nicola.spaldin@mat.ethz.ch}

\author{Sang-Wook Cheong}
\affiliation{Keck Center for Quantum Magnetism and Department of Physics and Astronomy, Rutgers University, Piscataway, New Jersey 08854, USA}

\author{Sin\'ead Griffin}
\affiliation{Materials Sciences Division, Lawrence Berkeley National Laboratory, Berkeley, California 94720, USA}
\affiliation{Molecular Foundry, Lawrence Berkeley National Laboratory, Berkeley, California 94720, USA}

\begin{abstract}
We review the conditions that cause or prohibit non-relativistic spin splitting of the energy bands in antiferromagnets. We propose that the existence of spin splitting in magnetically ordered systems is the default scenario and outline the criteria that must be met to avoid it. We discuss some of the properties of those special antiferromagnets that succeed in preserving their spin degeneracy. 
\end{abstract}

\maketitle

\section{Introduction}

There has been tremendous recent interest in antiferromagnetic materials that have an energy splitting of non-relativistic origin between bands of opposite spin orientation \cite{Hayami/Yanagi/Kusunose:2019,Yuan_et_al:2020,Yuan_et_al:2021,Smejkal/Sinova/Jungwirth:2022_1,Yuan/Zunger:2023}. Such antiferromagnets break global time-reversal symmetry in spite of their net zero magnetization and are associated with a plethora of related properties such as tunneling magnetoresistance \cite{Shao_et_al:2021}, anomalous Hall response \cite{Smejkal_et_al:2020} and surface magnetization \cite{Bhowal_et_al:2025}. They have even been given their own name, and are now affectionately referred to as {\it altermagnets} in the literature \cite{Smejkal/Sinova/Jungwirth:2022_2}. 

Given their intriguing properties, it is not surprising that much current activity is focused on expanding the zoology of candidate altermagnetic materials. Initially, suitable materials were identified largely by chemical intuition \cite{Yuan_et_al:2020}, until systematic guidelines based on a multipolar classification using the polar electric and magnetic toroidal multipoles were introduced \cite{Hayami/Yanagi/Kusunose:2020}. Design principles based on crystal chemical factors have subsequently been developed \cite{Wei_et_al:2024,Fender/Gonzalez/Bediako:2025}, and AI-accelerated discovery has been used to identify altermagnetic materials within databases \cite{Gao_et_al:2025}. 
 A handy computer code even exists to determine whether your favorite compensated antiferromagnet is altermagnetic or not given its magnetic and crystallographic symmetry \cite{10.21468/SciPostPhysCodeb.30,10.21468/SciPostPhysCodeb.30-r1.0}. 

In this work, we start from the alternative premise that altermagnetism is in fact a rather common state for a magnetic material to find itself in. As a result, a somewhat unusual set of circumstances must coincide for a magnetic material to {\it avoid} being altermagnetic. This is clear when one recognizes that in general, ordering of magnetic moments breaks time-reversal symmetry, which in turn splits the bands corresponding to opposite orientations of those moments. The only exception is when the time-reversal is combined with some other special symmetry operation that restores the Kramers degeneracy. 
Here, we review the requirements that allow magnetic orders to avoid having spin-split bands, show how most materials fail to meet these requirements, and describe the special properties of such non-altermagnetic antiferromagnets. We hope that our discussion restores fondness within the materials community for the class of materials that have become rather disparagingly known as ``conventional antiferromagnets.''

\section{Standard-Model Altermagnets}

The ``standard model'' of altermagnetism (SMALM) calls for collinear antiferromagnets in which the opposite spins occupy two distinct sublattices that are related by a symmetry, ${\cal C}$, that is neither inversion,  ${\cal P}$, nor a fractional lattice translation, ${\vec t}$. (We will refer to these symmetry-related lattices occupied by opposite spins as ``spin sublattices'' throughout this work). What is the role of each of these conditions? The antiferromagnetic order breaks time-reversal symmetry, ${\cal T}$. This is a necessary (but not sufficient) condition for the energy of an electron with spin orientation $\sigma$ at a particular point in $k$ space, $E_{k, \sigma}$, to be different from that of an electron with opposite spin orientation at the same $k$ point, $E_{k,-\sigma}$, in the absence of spin-orbit coupling. In turn, it is a requirement for non-relativistic spin-splitting. (When spin-orbit coupling is present, the absence of inversion symmetry allows spin-split bands even when time reversal is preserved, leading for example to the well-established Rashba and Dresselhaus effects). The symmetry relation, ${\cal C}$, between the opposite spin sublattices makes things nice and tidy by enforcing spin degeneracy at $\Gamma$ and giving the pattern of spin splitting an attractive shape, with symmetry-enforced sign changes and nodes, such as $d$-wave or $g$-wave. It also guarantees the zero magnetization criterion (again in the absence of spin-orbit coupling) even when orbital magnetism is included. Combined $\cal{P} \cal{T}$ symmetry, however, forces  an $E_{k,\sigma} = E_{k,-\sigma}$ degeneracy even when ${\cal T}$ is broken (see BOX 1) and so the ${\cal P}$  operation has to be excluded from ${\cal C}$. Likewise, for collinear spin configurations, the presence of a fractional lattice translation ${\vec t}$ connecting the two opposite spin sublattices also forces $E_{k,\sigma} = E_{k,-\sigma}$ degeneracy even when ${\cal T}$ is broken (see BOX  2), excluding ${\vec t}$ as an allowed symmetry. 

In fact, we already see the answer to the question posed in the title: For a compensated, ${\cal T}$-symmetry breaking antiferromagnet to {\it avoid} spin-splitting of its bands, it must have either ${\cal P \cal T}$ symmetry (BOX 1) or have its opposite spin sublattices related to each other by a fractional lattice translation, ${\vec t}$ (BOX 2).

\begin{boxA}
{\bf BOX 1. Effect of ${\cal P}$  and ${\cal T}$
symmetry operations on spins, momenta and band energies.}\\

The spin of an electron, $\sigma$, which transforms  like an angular momentum, changes sign under time reversal ${\cal T}$ but is unaffected by space inversion ${\cal P}$:
\begin{eqnarray*}
{\cal T} \sigma & = & -\sigma \\
{\cal P} \sigma & = & \sigma 
\end{eqnarray*}
In contrast, the momentum $k$, which transforms in the same way as a velocity, is reversed by both operations:
\begin{eqnarray*}
{\cal T} k & = & -k \\
{\cal P} k & = & -k 
\end{eqnarray*}
Then it's straightforward to see that the band energy at a particular wave number $k$ and spin $\sigma$,  $E_{k,\sigma}$,  transforms under time-reversal  and space-inversion operations as follows:
\begin{eqnarray*}
{\cal T} E_{k,\sigma} & = & E_{-k,-\sigma}\\
{\cal P} E_{k,\sigma} & = & E_{-k,\sigma} 
\end{eqnarray*}
and in turn
\begin{equation*}
    {\cal P} {\cal T} E_{k,\sigma} = E_{k,-\sigma} \quad.  
\end{equation*}

We see that spin splitting, with $E_{k,\sigma} \ne E_{k,-\sigma}$, requires breaking of ${\cal P} \cal{T}$ symmetry, which is achieved by breaking {\it either} $\cal{P}$ or $\cal{T}$ but not their product. Breaking of $\cal{P}$ gives rise to an antisymmetric spin splitting in momentum space as seen in the Rashba and Dresselhaus spin–orbit interactions, whereas breaking of $\cal{T}$ leads to symmetric spin splitting with respect to $k$, as seen in conventional ferromagnets and altermagnets. If both $\cal{P}$ or $\cal{T}$ are broken but their product $\cal{PT}$ remains a symmetry, the bands remain doubly degenerate.

\end{boxA}

\begin{boxA}
{\bf BOX 2. Global versus local time-reversal symmetry.}\\

All ordered spin systems break local time-reversal symmetry, since every spin changes sign under time reversal. In some collinear crystalline antiferromagnets, however, translation by a fractional lattice translation of the antiferromagnetic unit cell, ${\vec t}$, also changes the sign of each spin, that is
\begin{equation*}
{\vec t} \sigma  =  -\sigma \quad .
\end{equation*}
Since lattice translation does not affect the wavevector $k$, i.e.
\begin{equation*}
{\vec t} k  =  k \quad ,
\end{equation*}
we see that, for this class of antiferromagnets,
\begin{equation*}
{\vec t} E_{k,\sigma}  = E_{k,-\sigma} \quad ,
\end{equation*}
and there is no spin splitting of the bands in spite of the ${\cal T}$ symmetry breaking at the local level. In these special antiferromagnets, $\vec{t} \cal{T}$ remains a symmetry operation and the system is said to retain {\it global} time-reversal symmetry. Such global time-reversal symmetry can only occur when the magnetic order increases the size of the chemical unit cell by an integer multiple, and ${\vec t}$ is a lattice vector of the original chemical unit cell. 
\end{boxA}

\begin{figure}
\vspace*{11pt}
    \centering
    \includegraphics[width=1\linewidth]{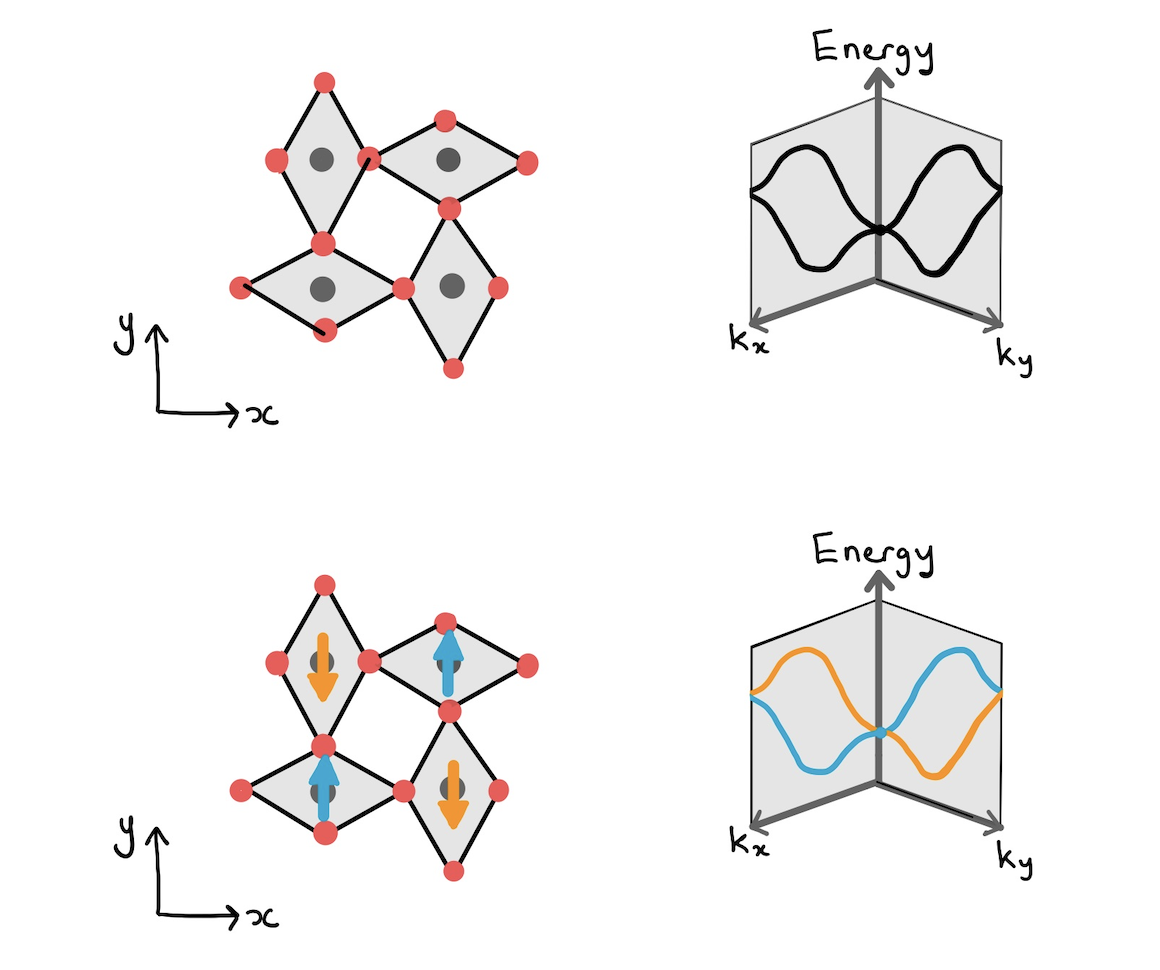}
    \caption{Upper panel: The crystal field from the ligand (red circles) polyhedra surrounding the metal ions (gray circles) splits the energies of the metal orbitals oriented in the $x$ and $y$ directions. Lower panel: When the pattern of antiferromagnetic spin ordering matches that of the structural motif, the split energy bands are decorated by the spin orientation. Note that the orientation of the opposite spins relative to the lattice, shown here as up and down, is arbitrary.}
    \label{Figure1}
\end{figure}
In practice, the SMALM requirements are met when a structural motif arranged in an alternating pattern is accompanied by an antiferroic ordering of spin dipoles, and the structural and spin orders share the same pattern of alternations in at least one direction \cite{Yuan/Zunger:2023}. The non-relativistic spin splitting (NRSS) can then be straightforwardly understood as follows: 1) The crystal field from the local structural distortion causes the bands formed from different atomic orbitals to have different energies. For example, in the top left polyhedron in Figure ~\ref{Figure1}, an electron in an orbital on the central atom oriented along $x$ has higher energy than that oriented along $y$, since it experiences greater Coulomb repulsion from the ligands. 2) The alternation of the local structural distortion causes the band splitting in adjacent structural motifs to have opposite orbital character. We see this by comparing the top right polyhedron, where the electron on the central atom oriented along $x$ is now in the lower energy level. For this particular pattern the bands have a ``$d$-wave'' symmetry, with identical orbital band splitting along $\pm k_x$ or $\pm k_y$, and opposite orbital band splitting between $k_x$ and $k_y$. 3) When the pattern of the spin dipolar order matches that of the orbital order, the split bands are dressed by the spin orientations giving the characteristic spin splitting of Figure~\ref{Figure1} (lower panel). 

This orbital -- spin pairing provides an ``effective spin-orbit interaction'' \cite{Hayami/Yanagi/Kusunose:2019} that is distinct from the usual {\it atomic} spin-orbit coupling. It can be convenient to recognize that the product of antiferroic spin dipoles and antiferroic orbitals arranged in a matching pattern is equivalent to a ferroically ordered magnetic multipole of higher than dipolar order \cite{Hayami/Yanagi/Kusunose:2020} and that the ferroic ordering of this higher-order magnetic multipole breaks global time-reversal symmetry. For the cartoon symmetry shown here, the ferroically ordered quantity is the magnetic octupole, which is a composite of the magnetic dipole and the charge quadrupole \cite{Bhowal/Spaldin:2024}.

Recognizing that the origin of SMALM is this paired antiferroic structural and magnetic ordering immediately explains why it is so common:
Many materials contain such an alternating structural motif. For example, when a local symmetry reduction occurs due to Jahn-Teller distortion, there is a strong preference for the polyhedra to align in a checkerboard pattern similar to the cartoon of Figure~\ref{Figure1}, since a ferroic alignment would introduce a large lattice strain. By the same reasoning, in systems containing bonds of different lengths, the long and short bonds tend to alternate rather than align; here the NiAs structure adopted by MnTe or the corundum structure of Fe$_2$O$_3$ are examples. Similarly, in three-dimensionally coordinated structures with corner-shared polyhedra, such as the perovskites, when a relative tilting of the polyhedra is present the polyhedra are forced to alternate their tilt orientations to maintain connectivity. The spin-dipole arrangement is in turn influenced by the pattern of structural order since the bond distances and angles set by the structural motif determine the exchange interactions between the magnetic dipoles. 
For example, if the structural motif has a checkerboard alternation, as is the case in the $a^-a^-a^-$ $R\bar{3}c$ perovskite structure, then all inter-motif transition metal bond angles and distances are the same, and correspondingly for isotropic electron configurations such as $d^3$ or $d^5$ all nearest-neighbor magnetic interactions are the same. If antiferromagnetic nearest-neighbor exchange dominates, then a G-type antiferromagnetic magnetic order with pattern of spin alternation identical to that of the structural alternation will result.
In a more complicated example, the common $Pbnm$ modification (known as the GdFeO$_3$ structure) of the cubic perovskite structure, has an $a^-a^-c^+$ pattern of octahedral rotations, which results in alternation of the polyhedra along all three crystallographic axes. As a result, for the commonly found A-type, C-type and G-type antiferromagnetic orders, there is always a spin alternation in the same direction as a structural alternation, and so all have an altermagnetic spin splitting, although along different directions in $k$ space \cite{Okugawa_et_al:2018}. We note that having the conditions for SMALMs``allowed by symmetry" does not guarantee a large splitting; in many systems the NRSS may be parametrically small, even if nonzero in principle.

\section{Beyond Standard-Model Altermagnets}

Having discussed why the criteria for satisfying standard-model altermagnetism are rather widely satisfied, we next discuss two popular extensions to the original definition that also yield a spin splitting of non-relativistic origin, and bring an even larger collection of antiferromagnets into the altermagnetic fold. First, in subsection~\ref{RelaxSymmetryConstraint}, we relax the constraint that the two opposite spin sublattices must be related by a symmetry, ${\cal C}$. Second, in subsection~\ref{RelaxCollinearity}, we allow spin non-collinearity. This can result from spin frustration, for example in triangular antiferromagnets that adopt a 120$^{\circ}$ spin structure, or, of course, from local atomic spin-orbit coupling, which can not be disabled in real materials. 

\subsection{Removing the constraint of the symmetry operation connecting opposite spins}
\label{RelaxSymmetryConstraint}

As described above, SMALMs have spin-structure motif pairs that are related to each other by spatial rotations or mirrors but not by translation or inversion. The symmetry relation enforces net zero magnetization in the absence of spin-orbit coupling even when orbital magnetization is considered and yields a distinctive momentum-dependent spin-splitting shape determined by the crystal symmetry. In this section, we relax the constraint that the spin-structure motif pairs must be related by a spatial rotation or mirror and explore two classes of beyond-standard-model altermagnets (BSMALMs): i) compensated antiferromagnets with chemically distinct magnetic atoms, and ii) compensated antiferromagnets with chemically identical magnetic atoms coordinated by different or differently arranged surrounding atoms. We emphasize that neither of these types of BSMALMs is new. The first includes the so-called half-metallic antiferromagnets that were first discussed, to our knowledge, in the 1990s \cite{Pickett:1998}, and the NRSS in the second was discussed in Refs. \cite{Yuan_et_al:2020,Yuan_et_al:2021}, where it was labeled spin-split type four (SST-4). 

We begin with the first type, in which the symmetry between the two opposite spin sublattices is broken by the magnetic sites having different chemistry. Usually, antiferromagnetically aligned inequivalent atoms will have a net uncompensated spin magnetic moment. In certain cases though, for example when two transition metal ions such as Fe$^{3+}$ and Mn$^{2+}$ differ by both the same atomic number and the same electronic charge, the spin magnetic moments are formally identical ($5 \mu_B$ in this case) and so formally compensate if they are anti-aligned. In addition, when at least one of the spin channels is insulating, the net spin magnetization is in general forced to be an integer number of Bohr magnetons, and in this case that integer is zero. The result is a practical, not only formal, zero magnetization in the absence of any orbital contribution. In fact, one can consider such a material to be a special case of a {\it ferri}magnet with a net spin magnetic moment of zero. The behavior can be realized in fully insulating systems, or, as mentioned above, in half metals, in which only one spin channel is insulating and the other is conducting \cite{Pickett:1998}. Such half-metallic antiferromagnets are particularly intriguing since they carry a fully spin-polarized current but have zero net magnetic moment.

We emphasize that, because the magnetic atoms are not related by any crystal symmetry, an antiunitary operation that enforces Kramers-like spin degeneracy does not exist for any $k$ point, and the spin degeneracy is lifted throughout reciprocal space. 
Such BSMALMs are $s$-wave altermagnets, in that their spin splitting is allowed to have the same sign throughout momentum space, in contrast to the alternating $d$- or $g$-wave spin-splitting pattern of conventional SMALMs. In this sense, half-metallic BSMALMs, with a splitting at $\Gamma$ and the possibility of having the same sign of spin polarization across the entire Fermi surface, are perhaps more technologically relevant than metallic SMALMs, in which the splitting at $\Gamma$  is zero and the sign of spin polarization depends on the orientation of the current. We also mention that one could classify such BSMALMs as {\it true} $s$-wave altermagnets, with ferromagnets classified as {\it false} $s$-wave altermagnets due to their residual magnetization. 

The second type, in which the symmetry between the two opposite spin sublattices is broken by the coordinating non-magnetic atoms, is illustrated in Ref. ~\cite{Yuan/Georgescu/Rondinelli:2024}, where altermagnetic nitride MnSiN$_2$ is chemically modified to break the two-fold rotation connecting the two magnetic Mn sites. On substituting half of the Si atoms with Sn, a finite NRSS at \( \Gamma \), which is precluded by Kramers'-type degeneracy in conventional SMALMs, develops, while perfect spin magnetic moment compensation is maintained.

\begin{figure}
    \centering
    \includegraphics[width=1\linewidth]{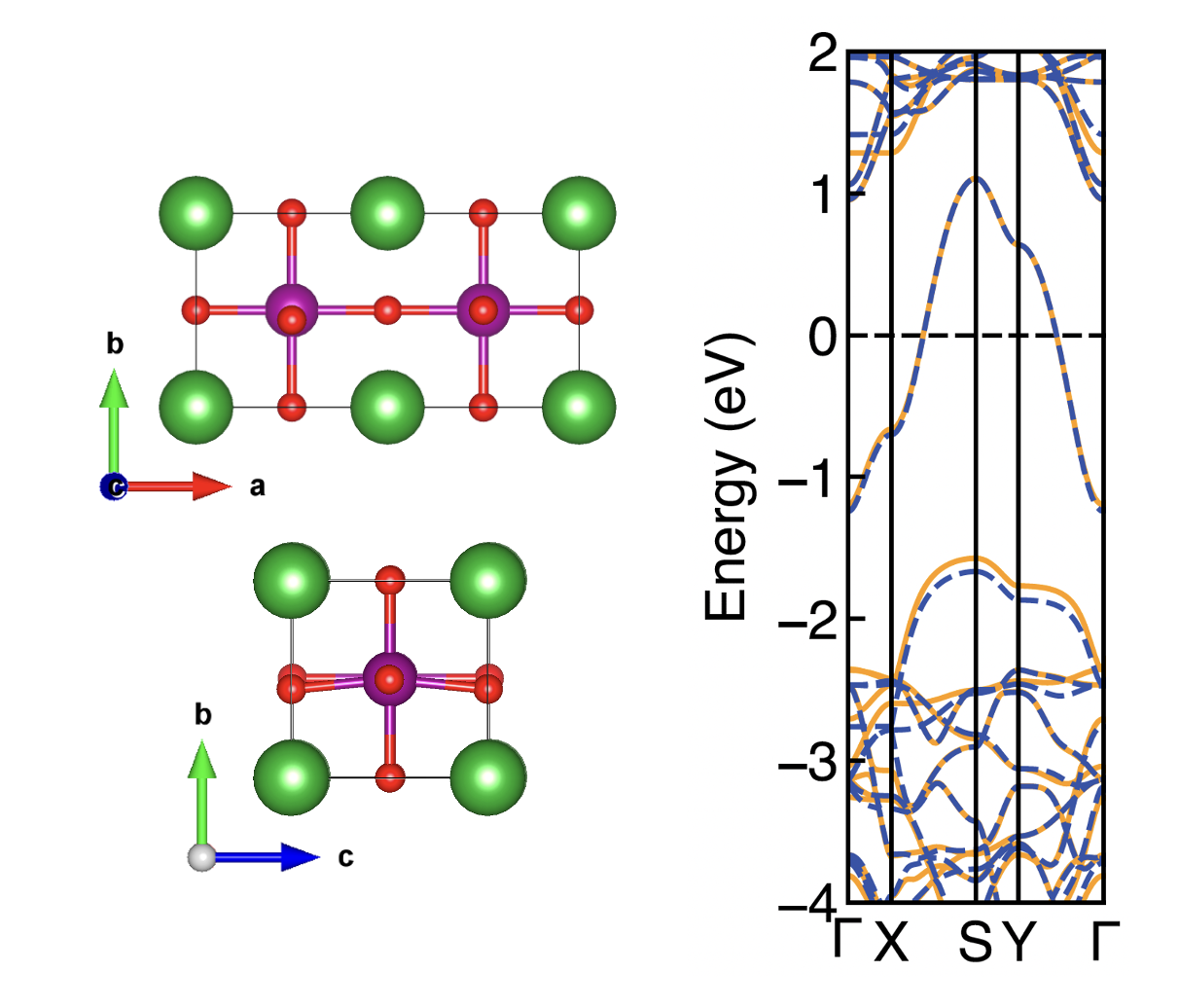}
    \caption{Left: Unit cell of cartoon LaMnO$_3$ consisting of two five-atom unit cells with oppositely spin polarized Mn ions and a small structural distortion introduced into one of the units. La ions are shown in green, Mn in purple and O in red. Right: The calculated band structure with the two spin channels shown in solid orange and dashed blue lines. Note in particular the spin splitting at the $\Gamma$ point.}
    \label{fig:LaMnO3}
\end{figure}

The same behavior results when the magnetic ions are surrounded by ions of the same chemistry but with different coordination polyhedra. We show a cartoon example in Figure~\ref{fig:LaMnO3}, where we take a doubled unit cell of cubic perovskite LaMnO$_3$ with antiferromagnetically aligned Mn ions and slightly distort the oxygens in the left polyhedron away from their ideal cubic positions. The resulting band structure, calculated using density functional theory within the VASP code \cite{Kresse/Furthmueller_PRB:1996} without including spin-orbit coupling, shows a spin splitting including at the $\Gamma$ point. 

In this spirit, perturbations that disrupt the symmetry relations between opposite-spin motifs need not be confined within a single crystallographic unit cell. The symmetry conditions required for NRSS can also be realized at the level of a larger superstructure, for instance in engineered, coupled magnetic–structural domain architectures. In particular, when a magnetic domain wall is locked to a structural twin boundary, the symmetry operations that enforce spin degeneracy in the bulk domains (such as $\cal PT$ or the translation-based ``global time-reversal'' symmetry) are generically absent for the full supercell, so that NRSS becomes symmetry-allowed in the supercell band structure. The resulting altermagnet-like, momentum-dependent spin polarization can exhibit symmetry-enforced nodal planes (when present) dictated by the magnetic space group of the full domain architecture)~\cite{Eggestad_et_al:2025}.

In summary, any perturbation that severs the symmetry link between opposite spin motifs, such as different chemistries of the magnetic ions or their coordinating ligands, unequal tilts of coordinating polyhedra, or even small polar displacements, will result in a BSMALM, provided that the magnetic ions have equal and opposite spin magnetic moments. Such systems can exhibit stronger, less symmetry‑constrained spin textures than SMALMs; in particular, the NRSS has no special symmetry, is not required to compensate throughout the Brillouin zone and can occur at the zone center.

\subsection{Allowing spin-orbit coupling and / or magnetic non-collinearity}
\label{RelaxCollinearity}

While the excitement about altermagnets rests on the non-relativistic origin of their spin splitting, in practice relativistic effects are always present in real materials. Spin-orbit coupling, even if it is small, can lead to a net magnetization even in formally compensated antiferromagnets due to spin canting (so called weak ferromagnetism), and / or a relativistic spin splitting of Rashba or Dresselhaus origin. These effects both cause non-collinear spin arrangements, the first in real space and the second in reciprocal space, and so are excluded in SMALMs. However, since spin-orbit coupling is unavoidable in real life, their existence and interaction with any accompanying NRSS are ignored at one's peril \cite{Cheong/Huang:2025}! Here we discuss some examples of BSMALMs in which an underlying NRSS can still be clearly identified, even when the spin-orbit coupling (SOC) results in non-collinear spins either in real or reciprocal space.

\begin{figure}
    \centering
    \includegraphics[width=\linewidth]{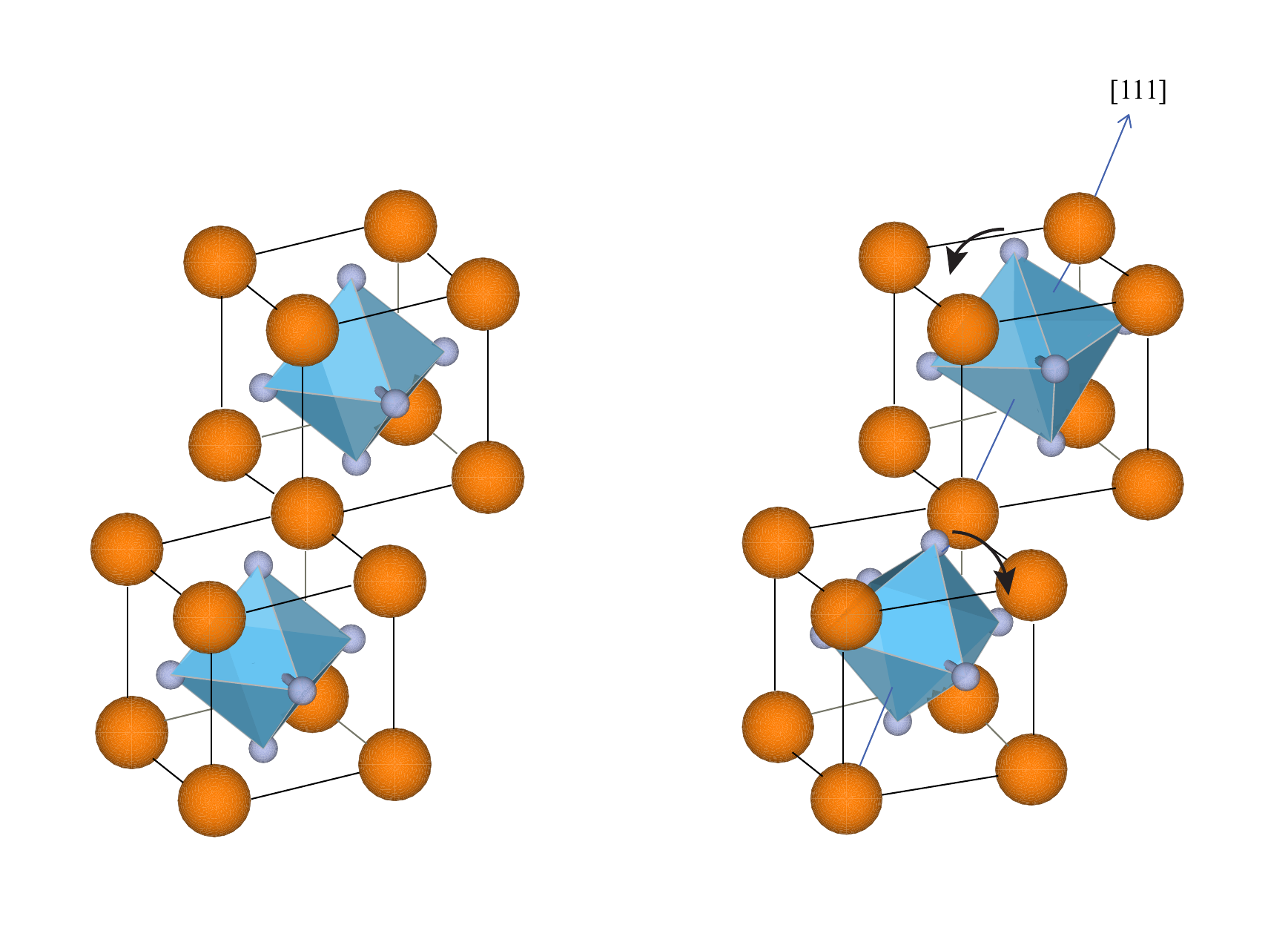}
    \caption{Crystal structure of ferroelectric BiFeO$_3$. The ground state $R3c$ crystal structure (right) is reached from the ideal cubic perovskite structure (left) via an antiferrodistortive rotation of the light blue FeO$_6$ octahedra (black arrows) around, and a displacement of the orange Bi ions along, the pseudocubic [111] direction.}
    \label{Figure3}
\end{figure}
As a prototypical example, we take multiferroic BiFeO$_3$ in its G-type antiferromagnetic phase. We note that while the original discussion of altermagnetism focused on centrosymmetric systems, multiferroics are, in general, promising altermagnetic candidates, since their ferroelectric polarization means that $\cal{P} \cal{T}$ symmetry is broken. The $R3c$ structural ground state of BiFeO$_3$ is reached from the ideal cubic perovskite structure via two distortions (Figure~\ref{Figure3}): a net displacement of the anions and cations along the [111] direction that yields the ferroelectric polarization and an antiferrodistortive rotation of the FeO$_6$ octahedra around the pseudocubic [111] axis. Since the magnetic anisotropy is easy plane, i.e. the Fe$^{3+}$ moments lie in the plane perpendicular to [111], there is a canting of the spins leading to a net weak ferromagnetic moment, so G-type BiFeO$_3$ does not qualify as a standard model altermagnet. A recent first-principles study \cite{Urru_et_al:2025} showed, however, that, in the absence of SOC, collinear G-type BiFeO$_3$ is a $g$-wave altermagnet with substantial NRSS away from the high symmetry lines of the Brillouin zone, and that the inclusion of SOC does not strongly perturb the altermagnetic character. In addition, the first-principles calculations confirmed the expectation from symmetry considerations, that the structural motif that enables the NRSS is the antiferrodistortive rotation of the oxygen octahedra. We note that the experimentally reported magnetic ground state of BiFeO$_3$ is in fact a long wavelength spin cycloid \cite{Sosnowska/Peterlin-Neumaier/Streichele:1982}, which was not included in this study. 

Interestingly, antiferromagnets with non-collinear spin spirals that break inversion symmetry can have a NRSS spin splitting that is odd / $p$ symmetry in $k$ \cite{Hirsch:1990}, with the spin polarization in the plane perpendicular to the plane of the spiral changing sign between $+k$ and $-k$ along the axis of propagation \cite{Hellenes_et_al:2024}. In insulating materials, such a spin spiral also causes an electric polarization, and as a result both the spiral orientation and the spin splitting can be controlled using an electric field \cite{Kimura_et_al_Nature:2003}; the electric-field control of spin splitting was recently demonstrated via this mechanism in NiI$_2$ \cite{Song_et_al:2025}. In metallic systems, non-reciprocal transport and anomalous Hall conductivity have been demonstrated \cite{Yamada_et_al:2025}. 

\section{So what is needed in order to be non-altermagnetic?}

Given that, with the extended definitions outlined above, there are myriad ways in which a material can be altermagnetic, in this final section we address the question of how a material can acquire non-altermagnetic status, and describe the special properties of such materials.

\subsection{First possibility: Not breaking time-reversal symmetry}

The first possibility is that a material does not break time-reversal symmetry at all, even locally. This can be achieved by the complete absence of local magnetic moments, for example in insulators containing only closed-shell atoms or ions; such a material would of course be conventionally described as non- (or dia-)magnetic. For the case of metals, the absence of local moments occurs in free-electron-like metals where Pauli physics (favoring two electrons of opposite spin occupying each orbital) dominate over Hund's exchange (favoring parallel spins) resulting in Pauli paramagnetism. Such materials are not antiferromagnetic in real space; if inversion symmetry is broken appropriately, spin-splitting is allowed in reciprocal space, and they can be considered to be $k$-space antiferromagnets. Paramagnetic materials with disordered local moments also retain time-reversal symmetry and so exclude altermagnetic behavior, although there has been a recent theoretical analysis of non-relativistic spin splitting in amorphous antiferromagnets which merits further study \cite{Dornellas_et_al:2025}.

\subsection{Second possibility: Breaking time-reversal symmetry but having an additional operation restore the Kramers degeneracy}

In order to be antiferromagnetic in real space, time-reversal symmetry must be broken at least locally. In this case, there are two options for avoiding altermagnetism -- either the product of time-reversal and space-inversion symmetries, $\cal{P T}$ must be a symmetry operation, or a spin flip followed by a fractional lattice translation that exchanges the atoms of the two spin sublattices, must restore the original system. We outline the special properties of non-altermagnetic antiferromagnets in both of these classes below.

\subsubsection{Antiferromagnets with $\cal{P T}$ symmetry}

The prototypical example of a material that breaks $\cal T$ and $\cal P$ symmetries, but not $\cal PT$, is corundum-structure chromia, Cr$_2$O$_3$. While the corundum structure has a center of inversion, the pattern of ordering of the magnetic moments below the N\'eel temperature breaks the $\cal P$ symmetry. Reversing the magnetic moments (i.e. applying time-reversal), followed by inverting the structure through the high-temperature inversion center (i.e. applying space inversion), restores the original arrangement. Only 21 of the 122 magnetic point groups ($\bar{1}'$, $2'/m$, $2/m'$, $m'mm$, $m'm'm'$, $4/m'$, $4'/m'$, $4/m'mm$,  $4'/m'm'm$, $4/m'm'm'$,  $\bar{3}'$, $\bar{3}'m$, $\bar{3}'m'$, $6'/m$, $6/m'$, $6/m'mm$, $6/mmm'$, $6/m'm'm'$, $m'\bar{3}'$, $m'\bar{3}'m$ and $m'\bar{3}'m'$) break $\cal T$  and $\cal P$ symmetry in the same manner such that $\cal PT$ is preserved. 

While Cr$_2$O$_3$ breaks both $\cal T$ and $\cal P$ symmetries, it is neither ferromagnetic nor ferroelectric, and its free energy, $F$, is quadratic in both electric and magnetic fields. Importantly, there is also a term that is bilinear in $\vec{E}$ and $\vec{H}$, so that
\begin{equation}
F\left(\vec{E},\vec{H}\right) = F_{0} - \frac{\varepsilon_{ij}E_{i}E_{j}}{8\pi} - \frac{\mu_{ij}H_{i}H_{j}}{8\pi} - \alpha_{ij}E_{i}H_{j} + ... \label{eq:Ftilde}
\end{equation}
where $\varepsilon_{ij}$, $\mu_{ij}$ and $\alpha_{ij}$ are the dielectric
permittivity, the magnetic permeability and the so-called {\it linear magnetoelectric} tensor respectively. Setting the partial derivatives of $F$ with respect to $\vec{E}$ ($\vec{H}$) equal to $-\vec{D}/4\pi$ ($-\vec{B}/4\pi$) yields
\begin{equation}
\begin{array}{rcl}
 P_{i} &=& \chi^{\rm e}_{ij} E_{j} + \alpha_{ij} H_{j} \\ \\
 M_{i} &=& \alpha_{ji} E_{j} + \chi^{\rm m}_{ij} H_{j}
\end{array}
\end{equation}
where $\chi^{\rm e}_{ij}$ and $\chi^{\rm m}_{ij}$ are the dielectric and magnetic susceptibility
tensors, respectively. We see that, in addition to the usual electric-field induced electric polarization and magnetic-field induced magnetization, such materials exhibit a linear magnetoelectric response, in which an applied electric (magnetic) field induces a magnetization (polarization) that is linear in the applied field. In the case of Cr$_2$O$_3$ the magnetic and crystallographic symmetry results in a diagonal linear magnetoelectric tensor, so that the induced magnetization (electric polarization) is parallel to the applied electric (magnetic) field.

All linear magnetoelectric materials are {\it ferromagnetoelectric}, in the sense that they contain local magnetoelectric multipoles on their constituent ions that are ferroically aligned, and lead to a net bulk magnetoelectric multipolization \cite{Spaldin_et_al:2013,Thole/Fechner/Spaldin:2016}, $\mathcal{M}_{ij}=\frac{1}{V}\int r_j m_i(\mathbf{r})d^3\mathbf{r}$, where $\vec{m} (\mathbf{r})$ is the magnetization density.
The $\mathcal{M}_{ij}$ tensor has the same symmetry as the magnetoelectric response tensor. For the case of Cr$_2$O$_3$, for which the $\mathcal{M}_{ij}$ and $\alpha_{ij}$ tensors are diagonal, the relevant ferroically ordered local multipoles are the magnetoelectric monopole, $a = \frac{1}{3} \int \vec{r} \! \cdot \vec{m}(\vec{r}) d^3 \vec{r}$, and the $z^2$ magnetic quadrupole,  
$q_{zz} = \int \left[r_z m_z - \frac{1}{3} \vec{r}\! \cdot \vec{m}(\vec{r}) \right] d^3\vec{r}$. Notably, linear magnetoelectric materials have a surface magnetization that is robust to roughness, resulting from a bulk-boundary correspondence with their magnetoelectric multipolization \cite{Spaldin:2021,Weber/Spaldin:2022,Weber_et_al:2024}.

\subsubsection{Antiferromagnets in which a lattice translation restores global time-reversal symmetry}

The prototype of our final class of non-altermagnetic antiferromagnets is the class of rock-salt structure transition-metal oxides such as NiO or MnO. While the chemical unit cell contains only two atoms (one transition metal and one oxygen), the antiferromagnetic checkerboard arrangement of the transition-metal magnetic dipoles generates a primitive magnetic cell containing two chemical formula units. As a result, time reversal followed by translation by a lattice vector of the chemical unit cell (which is a fractional lattice translation of the magnetic unit cell) restores the original configuration, global time-reversal symmetry is preserved, and the bands are doubly degenerate. We note that space-inversion symmetry is also preserved, both globally and at the atomic sites, and so relativistic spin splitting is also prohibited. 

We emphasize, that, although we commonly think of such materials as conventional AFMs, the criteria that must be fulfilled in order to achieve this scenario are stringent. The structure must be inversion symmetric, the magnetic dipoles must order antiferromagnetically in a manner that increases the size of the unit cell by an integer factor, and all other higher-order magnetic multipoles (since the Wyckoff positions are all centrosymmetric, these are the magnetic octupoles, the magnetic triakontadipoles, etc.) must also be antiferromagnetically ordered. A special feature of such non-altermagnetic antiferromagnets is that none of their surface planes have a roughness-robust magnetization \cite{Weber_et_al:2025}. 

\section{Summary}

In Summary, we have reviewed the conditions that cause or prohibit non-relativistic spin splitting of the energy bands in antiferromagnets. We argue that, since magnetic ordering breaks time-reversal symmetry, the existence of spin splitting is the default scenario. We outline the criteria that must be met to avoid it and discuss some of the special properties of non-NRSS antiferromagnets. We hope that this viewpoint unveils some of the mystery surrounding the newly classified altermagnets, and restores some enthusiasm for those antiferromagnets that manage to maintain their Kramers degeneracy.

\section*{Acknowledgements} NAS thanks the Miller Institute at UC Berkeley for sabbatical support and the Swiss National Science Foundation, project number TMAG-2\_225790 {\it Static and Dynamic Crystal Chirality} for funding.
SWC was supported by the DOE under Grant No.\ DOE: DE-FG02-07ER46382. SMG was supported by the U.S. Department of Energy, Office of Science, Office of Basic Energy Sciences, Materials Sciences and Engineering Division under Contract No.\ DE-AC02-05CH11231 within the Theory of Materials program.

\end{document}